\documentclass[conference]{IEEEtran}
\IEEEoverridecommandlockouts
\usepackage{cite}
\usepackage{amsmath,amssymb,amsfonts}
\usepackage{algorithmic}
\usepackage{graphicx}
\usepackage{textcomp}
\usepackage{xcolor}
\usepackage{multirow}
\def\BibTeX{{\rm B\kern-.05em{\sc i\kern-.025em b}\kern-.08em
    T\kern-.1667em\lower.7ex\hbox{E}\kern-.125emX}}
\newcommand{\squeezeup}{\vspace{-2.5mm}}
\begin{document}

\title{A Novel Frame Structure for Cloud-Based Audio-Visual Speech Enhancement in Multimodal Hearing-aids\\
}

\author{\IEEEauthorblockN{Abhijeet Bishnu\IEEEauthorrefmark{1}, Ankit Gupta\IEEEauthorrefmark{2}, Mandar Gogate\IEEEauthorrefmark{3}, Kia Dashtipour\IEEEauthorrefmark{3}, Ahsan Adeel\IEEEauthorrefmark{4}, Amir Hussain\IEEEauthorrefmark{3}, \\ Mathini Sellathurai\IEEEauthorrefmark{2}, and Tharmalingam Ratnarajah\IEEEauthorrefmark{1}}
\IEEEauthorblockA{\textit{ \IEEEauthorrefmark{1} School of Engineering}, \textit{University of Edinburgh}, Edinburgh, United Kingdom \\
Email: \{abishnu,t.ratnarajah\}@ed.ac.uk\\
\IEEEauthorrefmark{2} \textit{School of Engineering \& Physical Sciences}, \textit{Heriot-Watt Watt University}, Edinburgh, United Kingdom \\
Email: \{ankit.gupta,m.sellathurai\}@hwu.ac.uk \\
\IEEEauthorrefmark{3} \textit{School of Computing}, \textit{Edinburgh Napier University}, Edinburgh, United Kingdom \\ 
Email: \{m.gogate, k.dashtipour, a.hussain\}@napier.ac.uk \\
\IEEEauthorrefmark{4} \textit{School of Mathematics \& Computer Science}, \textit{University of Wolverhampton}, Wolverhampton, United Kingdom \\
Email: a.adeel@wlv.ac.uk}}

\maketitle
\begin{abstract}
In this paper, we design a first of its kind transceiver (PHY
layer) prototype for cloud-based audio-visual (AV) speech enhancement (SE) complying with high data rate and low latency requirements of future multimodal hearing assistive technology.
The innovative design needs to meet multiple challenging constraints including up/down
link communications, delay of transmission and signal processing, and real-time
AV SE models processing. The transceiver includes device detection, frame detection, frequency offset estimation, and channel estimation capabilities. We develop both uplink (hearing aid to the cloud) and downlink (cloud to hearing aid) frame structures based on the data rate and latency requirements. Due to the varying nature of uplink information (audio and lip-reading), the uplink channel supports multiple data rate frame structure, while the downlink
channel has a fixed data rate frame structure. In addition, we evaluate the latency of different PHY layer blocks of the transceiver for developed frame structures using LabVIEW NXG. This can be used with software defined radio (such as Universal Software Radio Peripheral) for real-time demonstration scenarios.
\end{abstract}


\section{Introduction}
Hearing impairment is one of the major public health issues affecting more than 20\% of the global population. Hearing aids are the most widely used devices to improve the intelligibility of speech in noise for the hearing impaired listeners. However, even the sophisticated hearing aids that use state-of-the-art multi-channel audio-only speech enhancement (SE) algorithms pose significant problems for the people with hearing loss as these listening devices often amplify sounds but do not restore speech intelligibility in busy social situations~\cite{r1}. In order to address the aforementioned issue, researchers have proposed audio-visual (AV) SE algorithms~\cite{r2}\cite{r3} that exploit the multi-modal nature of speech for more robust speech enhancement. In addition, various machine learning (ML) based SE have been proposed due to their ability to surpass conventional SE algorithms. These methods use the ML algorithm to reconstruct the clean audio signal from noisy audio signal. Some ML based algorithms include sparse coding~\cite{r4}, robust principal component analysis~\cite{r5}, visually derived wiener filter~\cite{adeel2019lip}\cite{adeel2020contextual}, non-negative matrix factorisation~\cite{r6} and deep neural networks~\cite{gogate2018dnn}\cite{gogate2020deep}. However,  despite significant research in the area of AV SE, deployment of real-time processing models on a small electronic device such as hearing aid remains a formidable technical challenge. Therefore, we propose a cloud-based framework that exploits the power of 5G and cloud processing infrastructure for real-time audio-visual speech enhancement. 

\begin{figure*}[!t]
\centering
\includegraphics[width=0.9\textwidth ,height=3in]{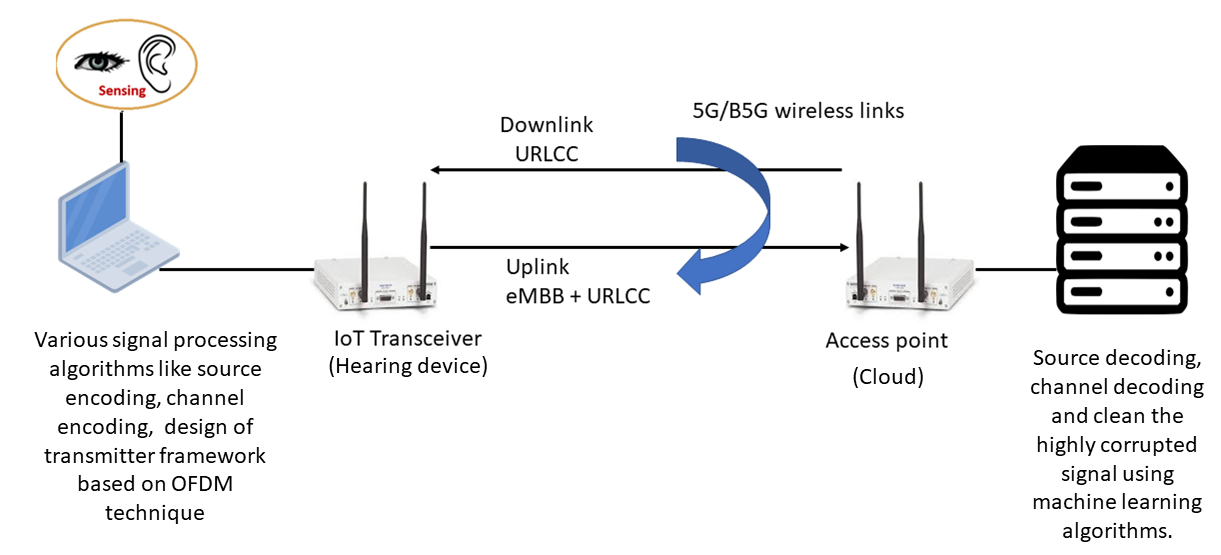}
\caption{Model of cloud-based audio-visual speech enhancement hearing aid}
\squeezeup
\squeezeup
\end{figure*}

We need a robust transceiver to send the raw or pre-processed data which includes AV signal from hearing aid device to cloud and get back clean signal from the cloud. The transceiver should meet the latency and data-rate requirements for synchronization of lip movement (visual) and the voice. In order to meet the stringent requirements of hearing aids, we developed both uplink (hearing aid to cloud) and downlink (cloud to hearing aid) transceiver, and in this paper our focus on physical layer of the proposed transceiver. Conventionally, downlink channel has high data rate as compared to uplink signal~\cite{r7}. However in hearing aids scenario, the uplink channel has high data rate due to transmission of AV signal, and the downlink has low data rate due to the transmission of only clean audio signal. Thus, uplink channel supports varying data rate as low data rate for audio signal only and high data rate for AV signal. The transmission and reception of data is based on frame which contains synchronization signal, reference signal, control channel and shared channel. Synchronization signal is used for timing and frequency synchronization, reference signal is used for wireless channel (between hearing aid and cloud) estimation, control channel is used for transmission of downlink or uplink control information, and the shared channel is used for downlink or uplink payload or data.

Rest of the paper is organised as follows. Section II describes the proposed model of cloud-based AV SE enhancement, source encoding and the proposed ML algorithm. Section III describes the downlink and uplink frame structure with proposed downlink and uplink control information. In section IV, latency of various proposed blocks of uplink and downlink frame structure is evaluated and finally conclude the paper in Section V.

\section{Model of Cloud based Hearing Aid}
Fig. 1 shows the proposed model of cloud-based AV speech enhancement hearing-aid. In the proposed model, the hearing aid device with small camera captures AV information and this information is compressed by source encoder. The compressed information is converted  into a frame structure for transmission over wireless channel to the cloud. In the receiver side (cloud), the received AV signal is decoded from the frame structure including source decoder. In order to clean the AV signal, machine learning algorithm is performed on the decoded AV signal and then the clean signal is transmitted from the cloud to the hearing aid device.

The next two subsection discuss the source encoder and ML algorithm for AV speech enhancement.

\subsection{Designing Low-Latency Audio Codecs}
The wireless transceiver design employs a source coding technique for efficient data compression-decompression. Thus, efficient source codes need to be designed for hearing-aid applications that map audio signals to the small binary codewords, also referred to as audio codecs. We can broadly classify the audio codecs as waveform codecs and parametric codecs. The waveform codecs aim to create codewords to obtain faithful reconstruction on a sample-by-sample basis by making no prior assumptions on the input audio signal. Thereby for general audio, waveform codecs produce very high-quality audio at medium-to-high bit rates but suffer from coding artefacts at low bit rates. The parametric codecs aim to create codewords to obtain perceptually similar reconstruction as the original audio by making strong prior assumptions on the input audio signal. Thus, parametric codecs perform well under low bit rates. Currently, OPUS~\cite{r8} and enhanced voice services (EVS)~\cite{r9} are the two most widely employed state-of-the-art audio codecs. The OPUS codec~\cite{r5} is standardized by the Internet Engineering Task Force (IETF) by including technologies from the Skype’s SILK codec and Xiph.Org’s CELT codec. The OPUS codec provides efficient compression for the transmission of interactive speech/music via the Internet, for example employed by YouTube. The EVS codec~\cite{r6} is standardized by the 3GPP by combining traditional coding tools such as LPC, MDCT and CELP. The EVS codec is a next-generation version of the AMR-WB mobile HD voice codec that provides the most efficient means for maintaining quality and efficiency in the wireless communication systems. Both the OPUS and EVS codecs provide high coding efficiency for varying audio signals, sampling, and bit rates, while enabling low-latency communications for audio signals in the real-time. We summarize the characteristics of the OPUS and EVS codecs in Table I.  

However, both OPUS and EVS codecs are not able to perform well for low bit rates. Thus, data-driven deep learning (DL) approaches have appeared as a promising solution for providing efficiently compressing and enhancing the audio signals. For example, a DL-based audio codec Sound-stream is recently proposed by Google~\cite{r10}. The Sound-stream audio codec is an end-to-end learning autoencoder-based framework that provides efficient audio signal compression at low bit-rates, while maintaining similar latency as EVS/OPUS. Although the latency of the state-of-the-art audio codecs (OPUS, EVS, Sound-stream) is low (around 30 ms), still the latency remains very large for hearing-aid scenario that targets a latency between 5-10 ms for end-to-end transmission, reception, and signal enhancement. Further, hearing-aid presents the challenge of operation under low bit-rates for real-world scenarios. Thus, there is an urgent need to develop audio codecs that can provide efficient audio compression and enhancement at low bit-rates, while maintaining the latency between 5-10 ms. 

\begin{table} [!t]

\begin{center}
\renewcommand*{\arraystretch}{1.25}
   \caption{Comparing state-of-the-art audio codecs} \label{tab:opus_evs}
    \squeezeup

    \squeezeup
    {\footnotesize
    
    \begin{tabular}{|c|c|c|} 
        \hline
        \textbf{Parameters} & \textbf{OPUS codec} & \textbf{EVS codec} \\
        \hline
        \hline
        Signal Bandwidth	& 4 kHz to 24 kHz	& 4 kHz to 20 kHz\\
        \hline
Supported Bit-rates	& 6 kbps to 510 kbps	& 5.9 kbps to 128 kbps\\
        \hline
Standardized By	& IETF (in 2012)	& 3GPP (in 2016)\\
        \hline
\multirow{2}{*}{Used By}	& YouTube, Skype,	& Voice over LTE\\
	& Zoom, MS Teams	& (VoLTE)\\
        \hline
Performance Comparison	 & \multicolumn{2}{c|}{EVS outperforms OPUS at low bit- rates}\\
        \hline
Latency Comparison	 & 26.5 ms &	32 ms\\
        \hline
    \end{tabular}}
  \end{center}
      \squeezeup
    \squeezeup

\end{table}

The hearing aid is a low power and complexity device. Thus, we need to have a low complexity transmitter in uplink (sending audio to the cloud) and a low complexity receiver in downlink (receiving audio at hearing-aid). Although AE-based Sound-Stream appears promising, AEs require neural networks on both the transmitter and receiver sides, thereby making it infeasible for our hearing-aid scenario. Thus, we propose to offload major processing in audio codecs to the cloud, which can have powerful GPUs for faster processing, thus reducing the latency. For example, we can apply model-based learning, where we utilize the OPUS/EVS audio codecs at the transmitter in uplink and apply DL-based audio decompression/enhancement.

\begin{figure}[!t]
\centering
\includegraphics[width=0.5\textwidth]{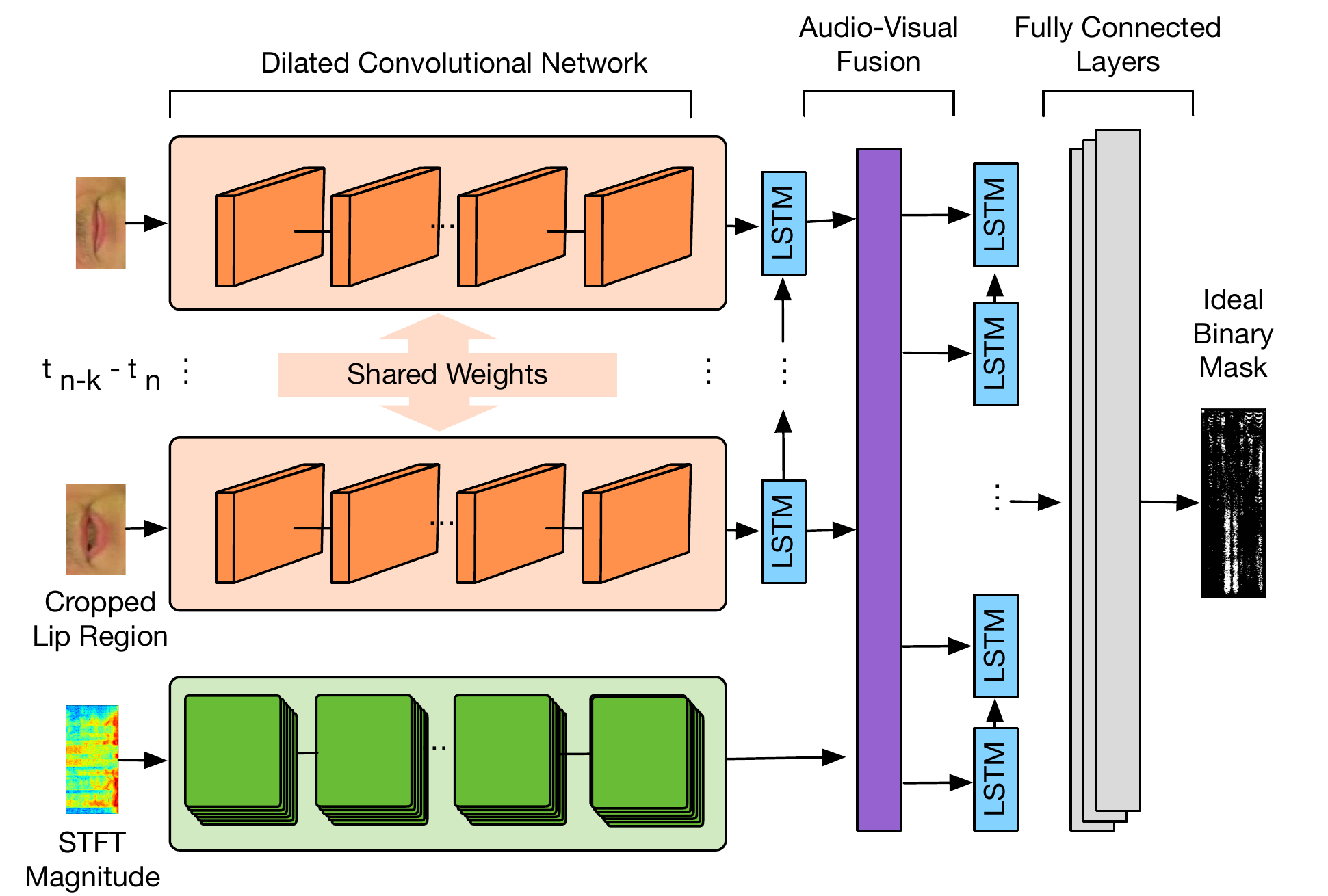}
\caption{Proposed AV Speech Enhancement model}
    \squeezeup
    \squeezeup
\label{fig:avse}
\end{figure}

\subsection{ML Algorithm of AV Speech Enhancement}
This section describes the AV SE algorithm used as part of the proposed framework. The model ingests cropped lip images of target speaker and noisy speech spectogram, as shown in Fig.~\ref{fig:avse}, to output a ideal binary mask that suppresses noise dominant regions and enhances speech dominant regions. Specifically, we adopted our previously proposed CochleaNet~\cite{r2} architecture to reduce processing latency by replacing dilated convolutions with depth wise separable convolutions, reducing the Short-time Fourier transform (STFT) window size from 78 ms to 32 ms, STFT window shift from 13 ms to 8 ms and by changing the number of convolutions in the audio feature extraction layers from 96 to 64. The visual feature extraction architecture is used without modification. The modified architecture is capable of processing streaming data frame-by-frame and can be used for real-time speech enhancement. The proposed model is evaluated using real noisy ASPIRE and VISION corpus~\cite{gogate2020visual}. 

\section{Frame Structure}
The downlink and uplink frame structure of hearing aid device is a grid type structure same as an LTE~\cite{r1}/5G-NR~\cite{r2} frame structure. The horizontal axis represents an orthogonal frequency division multiplexing (OFDM) symbol duration, while vertical axis represents an OFDM subcarriers. The duration of the frame is 10 ms which consists of 10 subframes of duration of 1 ms each. Each subframes consist of two slots of duration of 0.5 ms each and each slots contain seven OFDM symbols~\cite{r7}. The subcarrier spacing of 15 kHz is used for both downlink and uplink frame structure. The uplink channel (from hearing aid device to cloud) has varying data rate due to AV signal depends on the environment and hence uplink channel supports 1.4 MHz and 3 MHz bandwidth. However, the downlink channel (from cloud to hearing aid device) has constant data rate and hence it supports 1.4 MHz bandwidth. The parameters and their values for 1.4 MHz and 3 MHz are given in Table II. For 1.4 MHz bandwidth, the cyclic prefix (CP) length 10 is to be chosen for the  $0^{th}$ and $7^{th}$ OFDM symbol in each subframe and CP length 9 is to be chosen for the rest of the OFDM symbols in each subframe. Similarly for 3 MHz bandwidth,  CP length 20 is to be chosen for the  $0^{th}$ and $7^{th}$ OFDM symbol in each subframe and CP length 18 is to be chosen for the rest of the OFDM symbols in each subframe~\cite{r11}.

\begin{table}[!t]
\renewcommand{\arraystretch}{1.5}
\caption{Parameters value of different bandwidth}
    \squeezeup

\label{table_example}
\centering
\begin{tabular}{|c|c|c|}
\hline
\bf Parameters  & \bf  Values (1.4 MHz) & \bf Values (3 MHz) \\
\hline
Sampling Rate &  1.92 MHz & 3.84 MHz\\
\hline 
FFT Size & 128 & 256\\
\hline
\# of Data Subcarriers & 72 & 180\\
\hline
\# of Null Subcarriers  & 56 & 76\\
\hline
Cyclic Prefix  &  10 (9) &  20 (18)\\
\hline
\end{tabular}
    \squeezeup

\end{table}

\subsection{Downlink Frame Structure}
The downlink frame structure consists of downlink synchronization signal (DSS), physical downlink control channel (PDCCH), PDCCH reference signal (PDCCH-RS), downlink reference signal (DRS), and physical downlink shared channel (PDSCH). The DSS contains access point ID of cloud and used for frame detection. It is a 62 symbols (+1's and -1's)  long sequence which is same as the secondary synchronization signal of the LTE~\cite{r1} and mapped to the centre of an OFDM symbol. Two DSS is transmitted in a frame and it is mapped on the second last OFDM symbol of $0^{th}$ slot of $0^{th}$ and $5^{th}$ subframe of a grid.  The DRS are used to estimate the wireless channel between cloud and hearing aid device, and its generation is based on the access point ID of cloud, same as the generation of cell specific RS of an LTE~\cite{r7}. The DRS are  distributed in both time and frequency domain in a frame and their positions are based on the coherence bandwidth, coherence time and access point ID of cloud.  The PDCCH carries very crucial message block known as downlink control information (DCI) and it is transmitted by the access point to the active hearing aid devices. The DCI is of length 25 bits and its parameters are given in Table III.

\begin{table}[!t]
\renewcommand{\arraystretch}{1.5}
\caption{Bits of Downlink Control Information}
    \squeezeup

\label{table_example}
\centering
\begin{tabular}{|c|c|}
\hline
\bf MIB Parameters  & \bf  \# of Bits \\
\hline
Frame Number &  10\\
\hline 
Code Rate & 1\\
\hline
Modulation & 1\\
\hline
\# of Frames in a Transport Block & 4\\
\hline
End of Payload & 1\\
\hline
Uplink SS ID & 3\\
\hline
Reserved & 5 \\
\hline
\end{tabular}
    \squeezeup

\end{table}

In table III, ten bits are used to represent frame number. Two code rate is defined for downlink as 1/3 and 1/2 which are represented by bit 0 and 1 respectively. Similarly two modulation scheme is defined as quadrature phase shift keying (QPSK) and 16- quadrature amplitude modulation (16-QAM) which are represented by bit 0 and 1 respectively. Four bits are used for the number of frames in a transport block, one bit is used to represent the end of payload. Three bits are used to represent the uplink synchronization signal ID, which is used by the hearing aid device in conjunction with access point ID to generate unique SS for frame detection of hearing aid devices at the access point, and five bits are reserved for the future use. The DCI is carried by the PDCCH symbol and the generation of PDCCH symbols is shown in Fig. 3.  
\begin{figure}[!t]
\centering
\includegraphics[width=2.5in,height=2.5in]{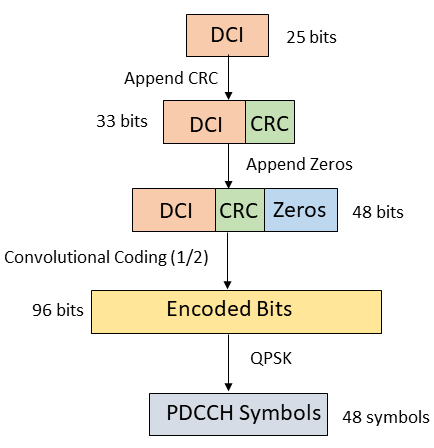}
\caption{PDCCH symbol generation.}
    \squeezeup
    \squeezeup

\end{figure}
The cyclic redundancy check (CRC) is calculated on the 25-bit DCI and append the 8-bit parity bits to the DCI. The parity bits are generated by the following generator polynomial~\cite{r12}:
\begin{equation}
g(D) = D^8 + D^7 + D^4 + D^3 + D + 1
\end{equation}
The CRC ensures the correct receiving of DCI. 15 zero bits (trace back length) is append to the 33 bits (DCI with parity bits) to form 48 bits, which is encoded by convolutional codes with code rate 1/2 and constraint length 3. The output of the convolutional codes is 96 bits which leads to 48 PDCCH QPSK modulated symbols. The PDCCH symbols are mapped on the 48 subcarrier of the last OFDM symbol of $0^{th}$ slot of $0^{th}$ and $5^{th}$ subframe of a grid. The PDCCH-RS is used to estimate the channel coefficients for the PDCCH symbol. The PDCCH-RS is generated using access point ID of cloud and mapped on the remaining 24 subcarrier of the last OFDM symbol of $0^{th}$ slot of $0^{th}$ and $5^{th}$ subframe of a grid. The remaining subcarriers of the grid is mapped with PDSCH symbol. The PDSCH symbol is generated  by encoding the payload by low-density parity check (LDPC) codes with code rate 1/3 and then modulated by either QPSK or 16-QAM.

\subsection{Uplink Frame Structure}
The uplink frame structure consists of uplink synchronization signal (USS), physical uplink control channel (PUCCH), PUCCH reference signal (PUCCH-RS), uplink reference signal (URS), and physical uplink shared channel (PUSCH). Irrespective of any bandwidth chosen, the generation and mapping of USS, PUCCH symbol, and PUCCH-RS are only based on the 1.4 MHZ. While, URS and PUSCH are based on the chosen bandwidth. The USS is used for frame detection. It is a 72 complex symbols  long sequence which is same as the sounding reference signal of the LTE~\cite{r7} and mapped to the centre of an OFDM symbol. It is generated with the help of USS ID and access point ID. Two USS is transmitted in a frame and it is mapped on the second last OFDM symbol of $0^{th}$ slot of $0^{th}$ and $5^{th}$ subframe of a grid.  The URS are used to estimate the wireless channel between hearing aid device and cloud, and its generation is based on the USS ID same as the generation of USS with access point ID set to zero. The URS are  distributed in both time and frequency domain in a frame and their positions are based on the coherence bandwidth, coherence time and USS ID.  The PUCCH carries very crucial message block known as uplink control information (UCI) and it is transmitted by the active hearing aid devices to cloud. The UCI is of length 25 bits and its parameters are given in Table IV.

\begin{table}[!t]
\renewcommand{\arraystretch}{1.5}
\caption{Bits of Uplink Control Information}
    \squeezeup

\label{table_example}
\centering
\begin{tabular}{|c|c|}
\hline
\bf MIB Parameters  & \bf  \# of Bits \\
\hline
Bandwidth & 1\\
\hline
Frame Number &  10\\
\hline 
Code Rate & 1\\
\hline
Modulation & 1\\
\hline
\# of Frames in a Transport Block & 4\\
\hline
End of Payload & 1\\
\hline
Reserved & 7\\
\hline
\end{tabular}
    \squeezeup
    \squeezeup

\end{table}

In Table IV, two bandwidth is defined for uplink as 1.4 MHz and 3 MHz which are represented by bit 0 and 1 respectively, ten bits are used to represent frame number, two code rate is defined for uplink as 1/3 and 1/2 which are represented by bit 0 and 1 respectively. Similarly two modulation scheme is defined as QPSK and 16-QAM which are represented by bit 0 and 1 respectively. Four bits are used for the number of frames in a transport block, one bit is used to represent the end of payload. Seven bits are reserved for the future use. The UCI is carried by the PUCCH symbol and the generation of PUCCH symbols is shown in Fig. 4.  
\begin{figure}[!t]
\centering
\includegraphics[width=2.5in,height=2.5in]{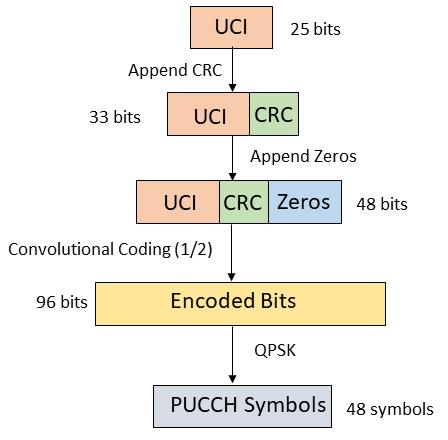}
\caption{PUCCH symbol generation.}
\squeezeup

\end{figure}
 The cyclic redundancy check (CRC) is calculated on the 25-bit DCI and append the 8-bit parity bits to the DCI. The parity bits are generated by the generator polynomial given in (1).
15 zero bits (trace back length) is append to the 33 bits (UCI with parity bits) to form 48 bits, which is encoded by convolutional codes with code rate 1/2 and constraint length 3. The output of the convolutional codes is 96 bits which leads to 48 PUCCH QPSK modulated symbols. The PUCCH symbols are mapped on the 48 subcarrier of the last OFDM symbol of $0^{th}$ slot of $0^{th}$ and $5^{th}$ subframe of a grid. The PUCCH-RS is used to estimate the channel coefficients for the PUCCH symbol. The PUCCH-RS is generated using USS ID and mapped on the remaining 24 subcarrier of the last OFDM symbol of $0^{th}$ slot of $0^{th}$ and $5^{th}$ subframe of a grid. The remaining subcarriers of the grid is mapped with PUSCH symbol. The PUSCH symbol is generated  by encoding the payload by low-density parity check (LDPC) codes with code rate 1/3 and then modulated by either QPSK or 16-QAM.

\begin{table}[!t]
\renewcommand{\arraystretch}{1.5}
\caption{Latency of various transmitter blocks of PHY layer of downlink signal}
\squeezeup
\label{table_example}
\centering
\begin{tabular}{|c|c|c|}
\hline
 \textbf{Blocks} &   \textbf{Time} &   \textbf{Occurrence}  \\
\hline
DSS & 10 $\mu$s & one\\
\hline 
DRS Symbol and Index  & 83 $\mu$s & one\\ 
\hline
 PDCCH Index & 15 $\mu$s & one\\ 
\hline
 Mapping & 12 $\mu$s & one\\ 
\hline
PDSCH Index & 4 ms & one\\ 
\hline
 Encryption & 20 ms & one\\
\hline 
RE2BIN &  2.3 ms &  Each frame (140 OFDM symbols) \\ 
\hline
LDPC Encoder & 4.9 ms &  Each frame\\ 
\hline
 Transport Block &  4.2 ms &  Each frame\\ 
\hline
PDCCH Symbol &  410 $\mu$s &  Each frame\\ 
\hline
OFDM Modulation  & 5.7 ms  & Each frame\\ 
\hline
Payload Generation & 18 ms  & Each frame\\ 
\hline
\end{tabular}
    \squeezeup
    \squeezeup

\end{table}

\begin{table}[!t]
\renewcommand{\arraystretch}{1.5}
\caption{Latency of various receiver blocks of PHY layer of downlink signal}
\squeezeup
\label{table_example}
\centering
\begin{tabular}{|c|c|c|}
\hline
 \textbf{Blocks} &   \textbf{Time} &   \textbf{Occurrence}  \\
\hline
DSS Det & 47 ms & one \\ 
\hline
 CFO Estimation \& Correction  &  152 $\mu$s &  Each frame\\ 
\hline
 PDCCH Decoder  &  805 $\mu$s  & Each frame\\ 
\hline
 Channel Estimation \& Correction &  5 ms &  Each frame\\ 
\hline
LDPC Decoder & 16.5 ms &  Each frame\\ 
\hline
BIN2RE & 932 $\mu$s & Each frame\\ 
\hline
PDSCH Symbol Detection & 22 ms  & Each frame\\
\hline
Decryption &  20 ms &  one\\
\hline 
\end{tabular}
    \squeezeup
    \squeezeup

\end{table}

\section{Results}
In this section, we have evaluated the latency of various physical layer blocks of transmitter and receiver of downlink and uplink frame. Latency of various transceiver blocks of PHY layer of transmitter and receiver is evaluated with Intel (R) Core (TM) i7-9700 CPU @ 3.00 GHz processor, 64 GB RAM, and LabVIEW NXG 3.0. The latency of various transmitter and receiver blocks of PHY layer of downlink frame are given in Table V and Table VI. For transmitter section of the downlink signal, the DSS symbol is generated once and then the same DSS is used for each frame, similarly, DRS symbol and its index are generated once and then the same symbol and index are used for each frame. The PDCCH index is generated for a single frame and then the same index is used for the remaining frame. In mapping, the DSS symbol and DRS symbol are mapped into a frame structure and the remaining positions or indices  (excluding DSS index, DRS index, and PDCCH index) of the frame are used for PDSCH data. The audio samples are encrypted by using Tangent-Delay Ellipse Reflecting Cavity-Map system and Non-linear Chaotic algorithm~\cite{r13}. The encryption occurs only once on the whole audio signal of size 46,440 samples. In RE2BIN, the audio samples are converted into single-precision floating point (32 bits for each sample) and this conversion performs for each frame (Each frame consists of 140 OFDM symbols of size either 128). The PDCCH symbol is generated in each frame due to the presence of system frame number which changes frame to frame. The transport block contains LDPC encoded PDSCH symbol, DSS symbol, DRS symbol and Convolutional encoded PDCCH symbol. The payload contains LDPC encoder, Transport block, OFDM modulation, PDCCH symbol generation and RE2BIN block for each frame.

In the receiver section of the downlink signal, the DSS Det is used to identify the access point ID and frame detection and it occurs only once at the beginning of the received signal. There is carrier frequency offset (CFO) due to carrier frequency mismatch between transceiver, hence CFO is to be estimated and then corrected for each frame. In PDCCH decoder, PDCCH symbol is decoded using PDCCH reference signal in each frame. Further processing of the received signal is based on the successful decoding (without any error) of the PDCCH symbol. In each frame, the wireless channel coefficients are estimated using DRS symbol which are used to decode the PDSCH symbol. The PDSCH symbol detection includes channel estimation and correction, and LDPC decoding in each frame. Then, the bits are converted into audio samples using BIN2RE block in each frame and finally decryption is to be performed on the encrypted received audio samples.

Similarly, the latency of various transmitter and receiver blocks of PHY layer of uplink frame are given in Table VII and Table VIII. 

\begin{table}[!t]
\renewcommand{\arraystretch}{1.5}
\caption{Latency of various transmitter blocks of PHY layer of uplink signal}
\label{table_example}
\centering
\begin{tabular}{|c|c|c|}
\hline
 \textbf{Blocks} &   \textbf{Time} &   \textbf{Occurrence}  \\
\hline
USS & 14 $\mu$s  & one\\ 
\hline
URS Symbol \& Index & 35 $\mu$s & one\\ 
\hline
PUCCH Index & 19 $\mu$s & one\\ 
\hline
 Mapping & 12 $\mu$s & one\\
\hline 
PUSCH Index & 3.5 ms & one\\
\hline
Encryption & 20 ms & one\\ 
\hline
PUCCH Symbol & 350 $\mu$s & Each frame\\ 
\hline
PUSCH Symbol Generation & 24 ms  & Each frame\\
\hline
\end{tabular}
\end{table}

\begin{table}[!t]
\renewcommand{\arraystretch}{1.5}
\caption{Latency of various transmitter blocks of PHY layer of uplink signal}
\label{table_example}
\centering
\begin{tabular}{|c|c|c|}
\hline
 \textbf{Blocks} &   \textbf{Time} &   \textbf{Occurrence}  \\
\hline
USS Det & 22 ms & one\\ 
\hline
PUCCH Decoder  &  750 $\mu$s & Each frame\\ 
\hline
Channel Estimation \& Correction & 8 ms & Each frame\\ 
\hline
PUSCH Symbol Detection &  30 ms & Each frame\\ 
\hline
Decryption & 20 ms & one\\ 
\hline
Data Enhancement  & 15 ms & one\\ 
\hline
\end{tabular}
\end{table}

\section{conclusion}
In this paper, we proposed the physical layer of a frame
structure for cloud-based audio-visual speech enhancement in future multimodal 
hearing aids. We have developed both uplink and downlink
frame structures to meet multiple constraints of high data-rate and low
latency. We also proposed various crucial control channels
within the frame structure which are required for reliable
communication. In addition, we conduct simulation experiments to evaluate the various
blocks of both uplink and downlink frame structures in terms
of latency. In future, we will further reduce the
latency of each block and implement the transceiver in field programmable
gate array technology for real-time latency evaluation.
\section{Acknowledgment}
This work is supported by the UK Engineering and Physical Sciences Research Council (EPSRC) programme grant: COG-MHEAR (Grant reference EP/T021063/1).

\bibliographystyle{IEEEtran}
\bibliography{ref}

\end{document}